\documentclass[aps,prl,twocolumn,superscriptaddress]{revtex4-1}

\usepackage{graphicx}
\usepackage{amsmath}
\usepackage{braket}
\usepackage{url}

\usepackage[colorlinks,
linkcolor=blue,
anchorcolor=blue,
urlcolor=blue,
citecolor=blue]{hyperref}

\begin{document}


\title{Two-neutron halo structure of $^{31}$F}


\author{N. Michel}\email[]{nicolas.michel@impcas.ac.cn}
\affiliation{Institute of Modern Physics, Chinese Academy of Sciences, Lanzhou 730000, China}
\affiliation{School of Nuclear Science and Technology, University of Chinese Academy of Sciences, Beijing 100049, China}
\author{J.G. Li}
\affiliation{School of Physics,  and   State Key  Laboratory  of  Nuclear  Physics   and  Technology, Peking University, Beijing  100871, China}
\author{F.R. Xu}
\affiliation{School of Physics,  and   State Key  Laboratory  of  Nuclear  Physics   and  Technology, Peking University, Beijing  100871, China}
\author{W. Zuo}\email[]{zuowei@impcas.ac.cn}
\affiliation{Institute of Modern Physics, Chinese Academy of Sciences, Lanzhou 730000, China}
\affiliation{School of Nuclear Science and Technology, University of Chinese Academy of Sciences, Beijing 100049, China}



\date{\today}

\begin{abstract}
We apply the Gamow shell model to study $^{25-31}$F isotopes. As both inter-nucleon correlations and continuum coupling are properly treated therein, the structure shape of $^{31}$F at large distance can be analyzed precisely. For this, one-nucleon densities, root-mean square radii and correlation densities are calculated in neutron-rich fluorine isotopes. It is then suggested that $^{31}$F exhibits a two-neutron halo structure, built from both continuum coupling and nucleon-nucleon correlations.
\end{abstract}


\maketitle

 \textit {Introduction.} --  Light exotic nuclei have been studied for several years using accelerators of last generation \cite{TANIHATA2013215,DOBACZEWSKI2007432}. Due to the use of radioactive ion beams, it has been possible to reach the neutron drip-line up to $^{40}$Mg \cite{Baumann20071022}. 
Contrary to well-bound nuclei, which are closed quantum systems, drip-line nuclei are open quantum systems, as they are either weakly bound or unbound with respect to particle emission. Furthermore, interesting phenomena appear at drip-lines, such as clusters in nuclei and halo structure \cite{PhysRevLett.112.162501}. Hence, the properties of drip-line nuclei should be studied to understand the nuclear force, which acts differently in the valley of stability and drip-lines.  

Several halo nuclei are known among $p$ shell nuclei, such as $^{6,8}$He \cite{PhysRevLett.93.142501,PhysRevLett.99.252501}, $^{11}$Be \cite{PhysRevLett.74.30} and $^{11}$Li \cite{PhysRevLett.55.2676}. A few halos are also known in $sd$ and $sdpf$ shell nuclei, such as the one-neutron halos of $^{31}$Ne \cite{PhysRevLett.112.142501} and $^{37}$Mg \cite{PhysRevLett.112.242501} and the two-neutron halo of $^{22}$C \cite{PhysRevLett.104.062701}. The principal ingredient of halos is an important occupation of the $s$ and $p$ continuum. Moreover, configuration mixing involving higher partial waves is also expected to be important for a proper description of halo nuclei. This is especially important for two-nucleon halo states. Theoretical calculations taking into account both these features are thus demanded to better understand halo structure.

The main models including both inter-nucleon correlations and continuum coupling in a complete fashion are the no-core shell model coupled with continuum (NCSMC) \cite{PhysRevLett.117.222501}, the Coupled-Cluster (CC) model \cite{PhysRevLett.108.242501} and the Gamow shell model (GSM) \cite{PhysRevLett.89.042501,PhysRevLett.89.042502,0954-3899-36-1-013101}. However, due to the huge model space dimensions, the NCSMC model could be used to describe drip-line nuclei of $A \sim 10$ nucleons at most \cite{Navr_til_2016}. Furthermore, only nuclei in the vicinity  of closed-shell systems can be treated within the CC model \cite{Hagen_2016}. Conversely, GSM has been used to calculate weakly bound and resonance states of the $p$, $sd$ and $pf$ shell nuclei \cite{PhysRevC.70.064313,PhysRevC.96.054316,PhysRevC.96.024308,PhysRevC.94.054302}. In particular, two-nucleon halos could be precisely described in the GSM framework \cite{PhysRevC.84.051304}. Consequently, GSM is an appropriate tool to precisely study  neutron-rich nuclei and halo structure.

Fluorine isotopes have been synthesized up to the neutron drip-line, which is reached with the loosely bound $^{31}$F isotope \cite{PhysRevLett.109.202503,wang2017ame2016}. They form a very interesting ground for theoretical studies, as they can provide with information about the proton-neutron interaction at drip-line. Moreover, the ground state of  $^{31}$F is suspected to be a neutron-halo state \cite{PhysRevC.64.011301}. Hence, both proton-neutron and neutron-neutron correlations in the continuum would be present in a nuclear system extended in space, which will surely rise to interesting phenomena. Thus, it is the object of the Rapid Communication to study weakly bound fluorine isotopes, and $^{31}$F in particular, with GSM.

This Rapid Communication is written in the following way. {}
The basic features of GSM are firstly quickly stated.
Then, the model space and Hamiltonians used in the calculation of fluorine isotopes will be presented. Afterwards, we will depict the ground state energies of fluorine isotopes at drip-line, as well as other observables of physical interest, such as one-nucleon density, root-mean-square radius and correlation density. They will allow in particular to reveal two-neutron halo structure in $^{31}$F. \\

{}

\textit {Model.} --  GSM is a configuration interaction approach based on the use of the Berggren basis \cite{BERGGREN1968265}. The Berggren basis possesses bound, resonance and scattering states, generated by a finite-range potential, such as a Woods-Saxon (WS) potential:
\begin{equation}
\sum_n \ket{n} \bra{n} + \int_{L^+} \ket{k} \bra{k}~dk = 1 \label{Berggren}
\end{equation}
where $n$ runs over bound and resonance states, and $L^+$
is a complex contour in the complex plane encompassing resonance states. The many-body basis used in GSM consists in the Slater determinants built from the Berggren basis of Eq.(\ref{Berggren}). Consequently, continuum coupling is included at basis level and inter-nucleon correlations are exactly taken into account via configuration mixing. GSM is then the tool of choice to study many-body halo and resonance states. Following the success of former GSM applications  \cite{PhysRevC.70.064313,PhysRevC.96.054316,PhysRevC.96.024308,PhysRevC.94.054302}, we will consider a model consisting of valence protons and neutrons interacting with a Furutani-Horiuchi-Tamagaki (FHT) interaction \cite{FHT1,FHT2} above an $^{24}$O core. The FHT interaction is a Gaussian-based residual interaction bearing central, spin-orbit and tensor terms, whose coupling constants are denoted as $V_c^{ST}$,$V_{\scriptscriptstyle{L\:\!\!S}}^{ST}$ and $V_{T}^{ST}$, respectively, where $S=0,1$ and $T=0,1$ are the spin and isospin of the two nucleons, respectively. It has been already used in the context of GSM to describe neutron-rich helium, lithium and beryllium isotopes, as well as radiative capture reactions in $A=6,7,8$ nuclei \cite{PhysRevC.96.054316,PhysRevC.91.034609,JPG_GX_Dong}. The $^{24}$O core is mimicked by a Woods-Saxon potential. 

The Hamiltonian is fitted to reproduce oxygen and fluorine isotopes at neutron drip-line. We included the ground state energies and a few excited states of $^{25,26}$O and $^{25-31}$F in the fit. As a consequence, the features of fluorine isotopes at drip-line can be properly assessed. Fluorine isotopes with fewer neutrons, closer to the valley of stability, are well bound, so that they do not need to be included in the GSM model space (see Ref. \cite{PhysRevC.93.051301} for a study of well bound fluorine isotopes with realistic interactions). The present formalism will also be used with an interaction issued from effective field theory (EFT), by fitting low energy parameters on experimental data, as done in Ref.\cite{PhysRevC.98.044301}. The latter approach will allow to compare two different Hamiltonians, on the one hand, and provide with a theoretical approach similar to the one used with realistic Hamiltonians \cite{SUN2017227}, on the other hand. 

As the single-valence proton is well bound in neutron-rich fluorine isotopes, it is sufficient to use the $0d_{5/2}$ and $1s_{1/2}$ harmonic oscillator (HO) basis states therein as proton valence states. The most important neutron partial waves, having a sizable coupling to the continuum, are $d_{3/2}$, $f_{7/2}$ and $p_{3/2}$. Indeed, considering a WS potential mimicking the $^{24}$O core, the $0d_{3/2}$, $0f_{7/2}$, $1p_{3/2}$ neutron states are close to particle-emission threshold and bear a neutron-emission width of about 1 MeV or smaller. We checked that the $1p_{1/2}$ and $0f_{5/2}$ neutron states have very large widths, of about 5 MeV and 9 MeV, respectively, so that associated partial waves can be neglected. This situation is similar to that occurring in the halo ground state of $^6$He, where the broad character of the $0p_{1/2}$ neutron state implies that $p_{1/2}$ contributions are negligible in the halo region \cite{PhysRevLett.89.042502,PhysRevC.84.051304}.

The Berggren basis is generated by a WS potential bearing loosely bound $0d_{3/2}$, $0f_{7/2}$ and $1p_{3/2}$ neutron one-body states, whose associated Berggren basis contours are complex. We also restrict the model space by demanding two occupied neutron states at most in the continuum. It is indeed sufficient to have a precision of less than 0.1 keV for widths. 
The values of the optimized WS parameters are the diffuseness $d$ = 0.65 fm, radius $R_0$ = 3.663 fm, the spin-orbit coupling $V_{\ell s}$ = 7.5 MeV fm$^2$, and central depth $V_0$ , which is equal to 65.659 MeV for protons (except for $\ell = 0$ using the EFT interaction, where it is 67.659 MeV) and, for neutrons, is equal to 39.978 MeV for $\ell = 0,2$, to 43.3 MeV for $\ell  = 1$ and to 39.9 MeV for $\ell  = 3$. The fitted parameters of the FHT interaction (see Tab.(\ref{Table.InterParamFHT})) differ substantially from the values obtained in Ref.\cite{PhysRevC.96.054316}. This mainly reflects their very different statistical uncertainties \cite{PhysRevC.96.054316}. For example, $V_c^{11}$, $V_{\scriptscriptstyle{L\:\!\!S}}^{11}$ and $V_{T}^{11}$, which are the most different in our calculation, exhibited the largest uncertainties in Ref.\cite{PhysRevC.96.054316}. Conversely, $V_c^{00}$ and $V_c^{10}$, which were found to be well constrained in Ref.\cite{PhysRevC.96.054316}, did not deviate much from their initial value. $V_c^{01}$ and $V_{T}^{10}$ varied significantly compared to their fitted values of Ref.\cite{PhysRevC.96.054316}, even though they were found not to be poorly constrained therein. However, their change is much less important than for the most sloppy parameters, bearing $S=1$ and $T=1$ (see Tab.(\ref{Table.InterParamFHT}) and Ref.\cite{PhysRevC.96.054316}). Hence, one can consider that the obtained values are consistent with those of
Ref.\cite{PhysRevC.96.054316}. 
\begin{table}[htb] 
\caption{\label{Table.InterParamFHT} Optimized parameters of the FHT interaction.
The parameters of the FHT interaction consist of central ($V_c^{ST}$), spin-orbit ($V_{\scriptscriptstyle{L\:\!\!S}}^{ST}$) and tensor ($V_T^{ST}$) parts (see Ref.\cite{PhysRevC.96.054316} for definitions). They depend on the spin $S=0,1$ and isospin $T=0,1$ of the two nucleons, respectively.}
\begin{tabular}{l|ccccccc}\hline \hline 
Parameter &$V_c^{11}$   & $V_c^{10}$    & $V_c^{00}$  & $V_c^{01}$    &$V_{\scriptscriptstyle{L\:\!\!S}}^{11}$   & $V_{T}^{11}$ & $V_{T}^{10}$  \\ \hline 
Value&  $-$59.9 &  $-$5.4  &  $-$24.3 &  $-$0.07 & 9.5 & 29.9 &  1.02 \\ \hline \hline
\end{tabular}
\end{table}
The fit of the parameters of the EFT interaction is shown on Tab.(\ref{Table.InterParamEFT}). The EFT parameters are separated in two parts: the first part is that of the leading order (LO) parameters, denoted as $C_S^{0,1}$, $C_T^{0,1}$, which are respectively spin-independent and spin-dependent, and where the isospin of the two nucleons, equal to $0,1$, explicitly appears; the second part consists of the next-to-leading order parameters, denoted as $C_{1 \dots 7}$ (see Ref.\cite{MACHLEIDT20111} for notation and definition of associated  operators). Note that $C_S^0$ and 
$C_T^0$ reduce to a single constant, so that we only fitted $C_S^0$ and arbitrarily put $C_T^0 = 0$. One can see that parameters are usually close to 1, as expected from the naturalness properties of low energy constants \cite{MACHLEIDT20111}. Indeed, all non-zero constants are situated between 0.1 and 3 in absolute value, with the sole exception of $C_2$, close to $10^{-3}$.
\begin{table}[htb]
\centering
\caption{\label{Table.InterParamEFT} Optimized parameters of the EFT interaction at leading order (LO) and next-to-leading order (NLO). They given in natural units.
The $C_S^{0,1},C_T^{0,1},C_{1 \dots 7}$ notations are taken from Ref.\cite{MACHLEIDT20111}. Parameters at leading order ($C_S^{0,1},C_T^{0,1}$) explicitly depend on the isospin $T=0,1$ of the two nucleons.}
\begin{tabular}{l|cccccccc} \hline \hline
LO constant &$C_S^{0}$   & $C_S^{1}$    & $C_T^{0}$  & $C_T^{1}$   \\  \hline 
LO value&  $-$0.12 &  $-$2.27  &  0 &  $-$0.73 \\ \hline  \hline 
NLO constant &$C_1$   & $C_2$    & $C_3$  & $C_4$  & $C_5$   & $C_6$  & $C_7$ \\ \hline 
NLO value&  0.20  &  0.001  &  0.25 &  0.10 & 0.25 & $-$0.52 & 0.17 \\ \hline \hline
\end{tabular}
\end{table}

\textit {Results.} --
\begin{figure}[!htb]
\includegraphics[width=1\columnwidth]{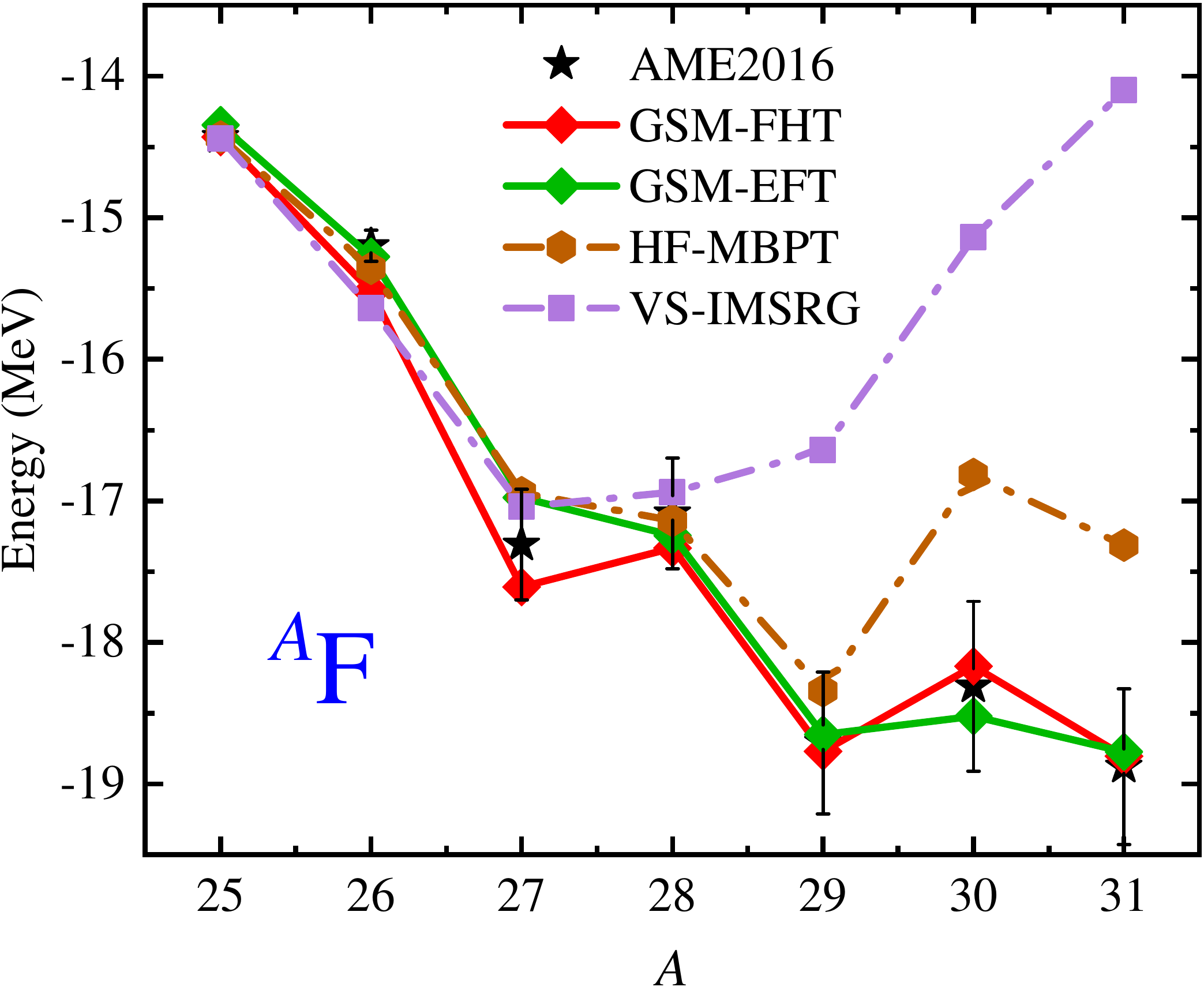}
\caption{Binding energies of $^{25-31}$F in MeV with respect to the $^{24}$O core calculated within different theoretical frameworks and compared to experimental data \cite{ensdf}. Besides the GSM calculations using FHT and EFT interactions of this work, calculations utilizing the HO basis, hence without continuum coupling, are presented with the HF-MBPT  and VS-IMSRG frameworks (color online).}\label{fig:1}
\end{figure}
The binding energies of fluorine isotopes are presented in Fig.(\ref{fig:1}).
Along with GSM calculations done with the FHT and EFT interactions and experimental data,
results issued from other theoretical calculations employing the HO basis, i.e.~many-body perturbation theory method (MPBT) based on a bound Hartree-Fock (HF) basis in an $sdpf$ cross-shell model space \cite{HJORTHJENSEN1995125,Wu_2019}, and the in-medium similarity renormalization group (IMSRG) method utilized in a $sd$ or $pf$ single-shell valence space (VS) \cite{PhysRevLett.118.032502,Holt:2019}, with bound HF basis states as well, are depicted. $^{25}$F has been fixed to its experimental energy in all used models on Fig.(\ref{fig:1}). One can immediately see that all methods reproduce the ground state energies of $^{25-28}$F isotopes, situated in the well-bound region, and start differing after $^{29}$F, hence when one reaches the neutron-drip line. This is particularly visible in VS-IMSRG, where the neglect of both multi-shell and continuum couplings at neutron drip-line generates a 4-5 MeV error in $^{30,31}$F. 
Conversely, the cross-shell couplings generated by the $sd$ and $pf$ shells are included in HF-MBPT .
Thus, HF-MBPT predicts proper binding energies up to $^{29}$F. Due to the lack of continuum coupling, however, the binding energies of $^{30,31}$F are about 1 MeV away from experimental error bars. One cannot make accurate predictions about a possible halo structure therein. 

On the contrary, GSM using a WS potential with FHT and EFT interactions correctly provide with binding energies up to $^{31}$F. Moreover, the odd-even staggering encountered from $^{28}$F, typical of the presence of a strong proton-neutron interaction, is well reproduced, with $^{30}$F being unbound and $^{31}$F being loosely bound. The $\chi^2$ deviation obtained with the FHT interaction is about 300 keV, which is comparable to the value of 250 keV of Ref.\cite{PhysRevC.96.054316}, whereas that provided by EFT is about 170 keV.  Note that the slightly different $\chi^2$ deviations obtained with the FHT and EFT interactions are in fact equivalent due to the large experimental error bars present in $^{29-31}$F, of the order of 1 MeV. We can then expect both interactions to provide with sensible observables other than energies, such as one-nucleon densities, root-mean-square (rms) radii and correlation densities, which we will consider in the following.
\begin{figure}[!htb]
\includegraphics[width=1\columnwidth]{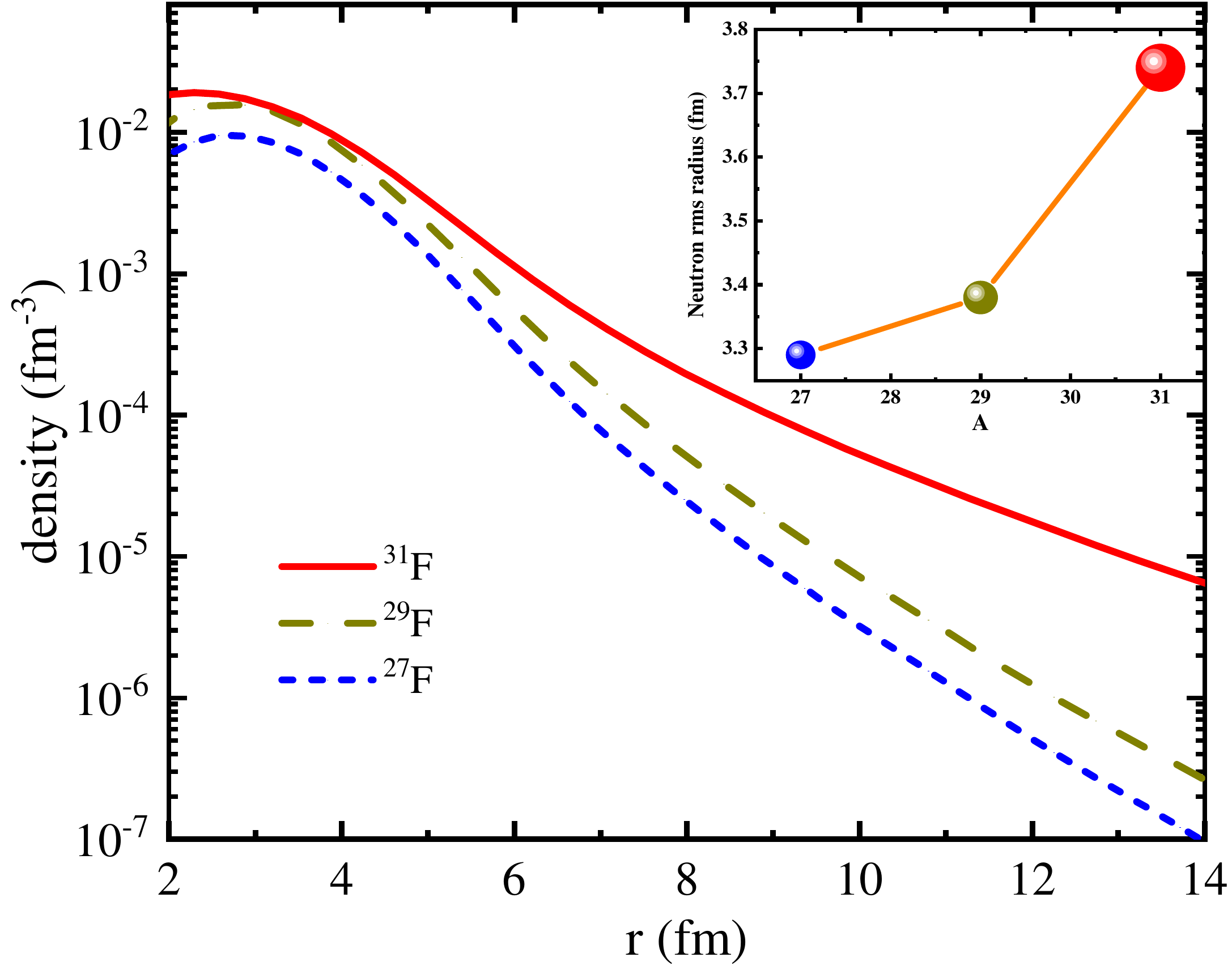}
\caption{One-nucleon densities (in fm$^{-3}$) of the bound $^{27,29,31}$F isotopes calculated with GSM using the EFT interaction in the valence space as a function of $r$ (in fm), respectively depicted by short-dashed, long-dashed and solid lines. The rms radii of these isotopes are shown in the insert (color online).}\label{fig:2}
\end{figure}

One will investigate the asymptote of the $^{31}$F ground state wave function. Its two-neutron separation energy is about 170 keV \cite{wang2017ame2016}, so that is sufficiently small to sustain a halo. It is indeed our assumption that the many-body wave function of $^{31}$F is principally made of a $^{29}$F sub-system and of two loosely bound valence neutrons mainly situated in the $p_{3/2}$ partial wave. Consequently, $^{31}$F would bear a two-neutron halo, similar to that present in $^6$He. In order to verify this assumption, we calculated the one-nucleon densities and neutron rms radii of the neutron-bound $^{27,29,31}$F isotopes with the EFT interaction (see Fig.(\ref{fig:2})). Indeed, a halo clearly develops in the asymptotic region of $^{31}$F. On the one hand, the one-nucleon density of $^{31}$F very slowly decreases on the real axis and is about one to two orders of magnitude larger than those of $^{27,29}$F in the asymptotic region. On the other hand, the neutron rms radius of  $^{31}$F does not follow the trend present in $^{27,29}$F as it sharply increases by about 0.4 fm, when the increase from $^{27}$F to $^{29}$F is about 0.1 fm. Results have been checked to be nearly identical when using the FHT interaction. 
\begin{figure}[!htb]
\includegraphics[width=1\columnwidth]{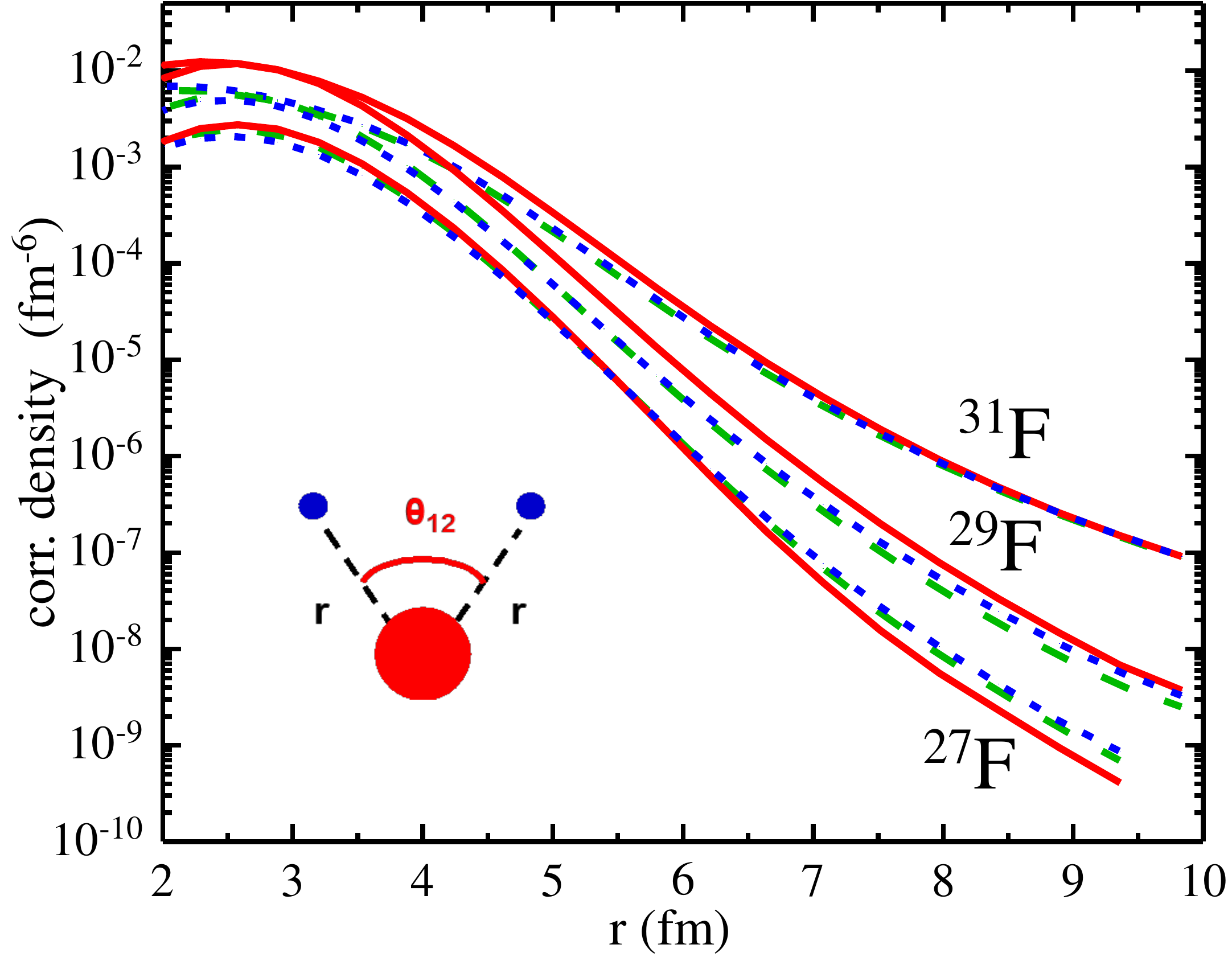}
\caption{Correlation densities (in fm$^{-6}$) of the bound $^{27,29,31}$F isotopes calculated with GSM using the EFT interaction as a function of $r$ (in fm).
The long-dashed, solid and short-dashed lines correspond to three values of the $\theta_{12}$ angle between the two nucleons, that is 32.75, 96.88, and 147.24 degrees, respectively (color online).}\label{fig:3}
\end{figure}

It is clear that $^{31}$F is very extended in space compared to $^{27,29}$F. One-nucleon density and neutron rms radius are nevertheless not sufficient to make definite statements about the possible two-neutron halo of $^{31}$F. Consequently, we calculated the correlation densities of $^{27,29,31}$F. The correlation density definition is standard \cite{PhysRevC.84.051304}:
\begin{equation}
\rho(r , \theta_{12}) = \braket{\Psi | \frac{\delta(r - r'_1)}{r r'_1} ~ \frac{\delta(r - r'_2)}{r r'_2} ~ \delta(\theta_{12} - \theta'_{12}) | \Psi} \label{correlation_density}
\end{equation}
where $r = r_1 = r_2$ is the distance between the core center of mass and the two nucleons, denoted as 1 and 2, $\theta_{12}$ is the angle between the two nucleons relatively to the core center of mass, and $\ket{\Psi}$ is the nuclear wave function, integrated over the space coordinates of the nucleons $r'_1$, $r'_2$ and $\theta'_{12}$. The Jacobian induced by the angular dependence is implicitly included in $\rho(r , \theta_{12})$,
similarly to the definition of Ref.\cite{PhysRevC.84.051304}. However, we added divisions by $r ~ r'_1$ and $r ~ r'_2$ in Eq.(\ref{correlation_density}) compared to the definition of Ref.\cite{PhysRevC.84.051304} to allow for a direct comparison with the one-body density. We show $\rho(r , \theta_{12})$ for three different $\theta_{12}$ angles, equal to 32.75, 96.88, and 147.24 degrees (see Fig.(\ref{fig:3})). We have checked that $\rho(r , \theta_{12})$ has a similar behavior for other angles at large distance. We can now analyze two-nucleon correlations in the wave function of $^{31}$F at large distance. Indeed, $\rho(r , \theta_{12})$ shows the same pattern as one-body density. While it falls off rapidly in the asymptotic region for $^{27,29}$F, its decrease for $^{31}$F is much slower, as for one-nucleon density (see Figs.(\ref{fig:2},\ref{fig:3})). As the proton part of the $^{31}$F wave function is very localized, the asymptotic regions are mainly generated from two delocalized neutrons. The correlation densities calculated with the EFT and FHT interactions have been checked to be qualitatively similar.
Thus, the asymptotic wave function of $^{31}$F is dominated by a halo of two neutrons, situated above a $^{29}$F well bound core. Moreover, $p_{3/2}$ one-body states are almost always present in the neutron configurations of the wave function of $^{31}$F, where configurations containing both $p_{3/2}$ and $d_{3/2},f_{7/2}$ neutron one-body states play an important role. Thus, all configurations contribute to build a complex two-neutron halo in $^{31}$F. \\

\textit {Conclusion.} -- Nuclei at drip-lines exhibit unique phenomena arising from the proximity of the continuum region and inter-nucleon correlations. One of the most important of them is halo structure, where one or two nucleons extends very far away from the nuclear region. It was the object of this Rapid Communication to demonstrate the two-neutron halo character of $^{31}$F. Indeed, besides its small two-neutron separation energy, it is an odd-even nucleus, so that both proton-neutron and neutron-neutron interactions participate to halo formation in $^{31}$F. By fitting the neutron-rich oxygen and fluorine isotopes with effective Hamiltonians, it has been possible to generate many-body wave functions recapturing the essential features of $^{25-31}$F. Consequently, the shape of $^{31}$F at large distance could be investigated by considering rms radius, one-nucleon density and correlation density, which clearly revealed a two-neutron halo structure in the wave function of $^{31}$F. This phenomenon has been noticed using two different effective interactions. As a consequence, it is very likely that $^{31}$F is a two-neutron halo nucleus. The study of nuclei at drip-lines in the $sdpf$ region is thus expected to present unique features where the radial extension of many-body wave functions and interactions between both protons and neutrons are intertwined. \\

\textit{Acknowledgments} -- This work has been supported by the National Natural Science Foundation of China under Grants No. 11435014, No. 11835001, No. 11575007 and No. 11847203;
the National Key R\&D Program of China under Grant No. 2018YFA0404401; 
 and the CUSTIPEN (China-U.S. Theory Institute for Physics with Exotic Nuclei) funded by the U.S. Department of Energy, Office of Science under Grant No. de-sc0009971. We acknowledge the High-Performance Computing Platform of Peking University for providing computational resources.

\bibliography{Fluorine_isotopes}

\end{document}